\documentclass[10pt]{article}    

\usepackage{amssymb,latexsym} 

\usepackage{amsbsy}  

\usepackage[leqno]{amsmath} 

\usepackage[mathscr]{eucal}  

\newfont{\sffl}{msbm10 at 16pt} 
\newfont{\sff}{msbm10 at 10pt}

\begin{document}           

\title{\vskip -.95in A Dirichlet-integral based dual-access collocation-kernel\\
        approach to point-source gravity-field modeling\thanks{\small{Officially approved for public  release; distribution is unlimited, but subject to both U.S. Government and arXive intellectual property rights protection.}}
}


\author{Alan Rufty\\         
\\
Naval Surface Warefare Center Dahlgren Division\\
 17320  Dahlgren Road\\
Dahlgren, VA. 22448--5100}
\maketitle                 

\newcommand{\KD}{K_{\text{D}}}
\begin{abstract}
 Problems in $\mathbb{R}^3$ are addressed where the scalar potential of an associated vector field satisfies Laplace's equation in some unbounded external region and is to be approximated by unknown (point) sources contained in the complimentary subregion.  Two specific field geometries are considered: $\mathbb{R}^3$ half-space and the exterior of an $\mathbb{R}^3$ sphere, which are the two standard settings for geophysical and geoexploration gravitational problems.  For these geometries it is shown that a new type of kernel space exists, which is labeled a Dirichlet-integral dual-access collocation-kernel space (DIDACKS) and that is well suited for many applications.  The DIDACKS examples studied are related to reproducing kernel Hilbert spaces and they have a replicating kernel (as opposed to a reproducing kernel) that has the ubiquitous form of the inverse of the distance between a field point and a corresponding source point.  Underpinning this approach are three basic mathematical relationships of general interest.  Two of these relationships---corresponding to the two geometries---yield exact closed-form inner products and thus exact linear equation sets for the corresponding point source strengths of various types (i.e., point mass, point dipole and/or point quadrupole sets) at specified source locations.  The given field is reconstructed not only in a point collocation sense, but also in a (weighted) field-energy error-minimization sense. 
\end{abstract}

\newcommand{\SubSec}[1]{

\vskip .18in
\noindent
\underline{{#1}}
\vskip .08in

}

\newcommand{\ls}{\vphantom{\big)}} 
\newcommand{\lsm}{\!\vphantom{\big)}} 

\newcommand{\mLarge}[1]{\text{\begin{Large} $#1$ \end{Large}}}
\newcommand{\mSmall}[1]{\text{\begin{footnotesize} $#1$ \end{footnotesize}}}

\newcommand{\Blbrac}{\rlap{\bigg{\lceil}}\bigg{\lfloor}} 
\newcommand{\Brbrac}{\bigg{\rbrack}} 

\newcommand{\R}[1]{${\mbox{\sff R}}^{#1}$} 
\newcommand{\mR}[1]{{\mbox{\sff R}}^{#1}}
\newcommand{\RR}{${\mbox{\sff R}}$}
\newcommand{\mRR}{{\mbox{\sff R}}}
\newcommand{\C}{${\mbox{\sff C}}$}
\newcommand{\mC}{{\mbox{\sff C}}}
\newcommand{\riii}{$\mbox{\sff R}^3$} 
\newcommand{\mriii}{\mbox{\sff R}^3}

\newcommand{\Dr}{\mathscr{D}_r}

\newcommand{\sps}{0} 
\newcommand{\spss}{0}

\newcommand{\smallindex}[1]{\text{\raisebox {1.5pt} {${}_{#1}$}}}
\newcommand{\eq}{:=}  

\vskip .05in
\noindent
\begin{itemize}
\item[\ \ ] \small{\textbf{Key words.} {Laplace's equation, inverse problem, Dirichlet form, point collocation, reproducing\\ \phantom{Key words. L} kernels,} 
fundamental solutions, point sources, multipole, potential theory}
\item[\ \ ] \small{\textbf{AMS subject classifications.} {35J05, 31B10, 86A22, 65D05}}
\end{itemize}

\renewcommand {\baselinestretch}{1.19} 

\section{Introduction}\label{S:intro}

 The goal of this article is to set forth a mathematical framework for the approximation of \riii\ harmonic fields in unbounded domains by point sources contained inside the complimentary region.  The proposed Dirichlet-integral dual-access collocation-kernel (DIDACK) approach has a mathematically and physically well-motivated underpinning.  The associated space (DIDACKS) has certain similarities to reproducing kernel Hilbert space (RKHS), but is distinct from it.  Two concrete \riii\ geometries are considered: (A) The harmonic field region consisting of half-space (denoted $\Omega_{1}$).  (B) The harmonic field region consisting of the exterior of a sphere (denoted $\Omega_{\sps}$).  Within this geometric context, the developed formalism easily handles various combinations of diverse types of point sources (such as point masses, point mass dipoles or point mass quadrupoles); moreover, for a set of specified source locations the formalism yields closed-form linear equation sets that simultaneously minimize the volume integrals of (weighted) field energy densities [i.e., (weighted) Dirichlet integrals]. 
 
   Techniques introduced here can be either applied directly or adapted for use in many mathematical and physical areas.  Examples on the mathematical side include potential theory, point source elliptic boundary value modeling (i.e., method of fundamental solutions), fast multipole techniques, radial-basis function techniques, RKHS techniques, geophysical collocation (GC) techniques and standard energy minimization based techniques (such as Galerkin and Raleigh-Ritz based approaches). Also see \cite{SandW}.  Examples on the physical side include geoexploration (where gravity is often used to locate oil or other minerals), geophysics,  magnetostatics (for a survey of these three areas see \cite{Kearey,Telford,Sleep,Blakely}), general electromagnetic source analysis, electrostatics and hydrostatics (for the physical significance of these three areas, and of field energy with regards to them, see \cite{Bergman}).  Also somewhat similar mathematical structures arise in many other areas such as  biophysical or biomedical engineering \{where, for example, electric dipole models are used in  electrocardiographical (ECG) modeling and current dipole models are used in electroencephalography (EEG)~\cite{Badia}\}.

   First, given the diversity just noted, it is clearly desirable to not confine potential readership to one specialty or another; hence, the required mathematical background for the core part of the article is limited to  classical \riii\ potential theory \cite{Kellogg,Jackson} and a basic understanding of functional analysis, as utilized in approximation theory.  This core part is primarily mathematical in content and makes up the first four sections.   Another, perhaps stronger, motivation exists for limiting the level of this core material---there is often a given mathematical level that certain research results tend to emerge from and the same results may not be apparent at either a higher or lower level of mathematical sophistication.  The first four sections, as well as the geophysical applications setting given in Section~\ref{S:PMhist}, are aimed at this default level of sophistication.   While a knowledge of neither RKHS nor GC is a necessity in these first \ref{S:PMhist} sections, a better understanding of both these areas is presupposed in Section \ref{S:GCconnections}, which describes the connections of DIDACKS theory to other kernel based approaches.  GC, which is often called least-squares collocation, is a RKHS based approach that differs from standard collocation techniques utilized by applied mathematicians in several relevant ways and, as such, may not be familiar to many readers.  Moritz \cite{Moritz} provides a readily accessible treatment of both RKHS and GC theory; nevertheless, the thrust of Section~\ref{S:GCconnections} should be directly accessible to those familiar with only standard RKHS theory \{\cite{Mate}, \cite{Aronszajn} or \cite{RKHSbook} (which contains \cite{Aronszajn})\}.

   Next consider the applications setting and why it was chosen.  While diverse candidate application areas exist, the approach adopted here is to select one particular representative problem area, namely geophysical gravitation, and discuss it thoroughly.  In addition to Laplacian inverse source theory, geophysical gravitation has two separate problem categories that point sources can be used in: field modeling and field estimation.  For gravitational modeling problems the field is assumed to be known throughout the region of interest and a more compact, but accurate, representation is desired.  For estimation problems it is assumed that values are known accurately at a certain number of points in the field region and that one wants to predict gravity values over some part of this field region.  Besides combinations of point sources, other techniques (such as GC) exist for treating gravitational field modeling and field estimation problems.  In a wider mathematical context, since the kernels involved satisfy a generalized collocation property, DIDACKS modeling and estimation approaches described here can be considered harmonic interpolation and extrapolation techniques, respectively.  With minor differences these gravitation ($\vec{G}$) problems use the same standard notation and techniques employed in electrostatics, where $\vec{E}$ is the field \cite{Jackson}, and so this arena should be readily accessible to all interested applied mathematicians, physicist and engineers.  (Within this electrostatic context the formalism developed here handles multiple types of various point sources---such as point charges, electrostatic dipoles or electrostatic quadrupoles.)   For consistency the notation used in the first four sections is also specialized to the gravitational setting, but in these sections the intrusion of this setting is minimal.

  Geophysical gravitation as an applications setting was chosen for the following five reasons:  (a) It has a strong overall historical association with potential theory.  (b) It has an easily understood notation and a readily accessible mature literature.  (c) It provides a family of  significant and challenging problems (that includes representative geophysical inverse source problems).  (d) DIDACKS theory was developed to handle gravitational problems and all of the author's direct numerical experience with it is in this arena.  (e) GC and its extensions are the family of approaches that are mathematically closest to DIDACKS theory.
 A discussion of point (a) is well outside the scope of the present article, but it is worth noting in this context that Gauss played a very significant role in the history of geophysics \cite[p. 1]{WandM}.  With regards to point (b), as well (c), see \cite{WandM,Moritz}.  (To gauge the advance of physical geodesy compare and contrast \cite{WandM} with \cite{HandM}, upon which it is based.)  With regards to (c) and (d), since no previous geophysical background is presupposed in the first four sections of the paper, the relevant overall geophysical and point mass modeling context is supplied in Section~\ref{S:PMhist}.  Since GC is the most commonly used approach for regional gravitational data processing and estimation it is briefly described at the end of Section~\ref{S:PMhist}.  The pertinent differences between GC and standard collocation techniques utilized by applied mathematicians are also described in Section~\ref{S:PMhist}.  Finally, Section \ref{S:GCconnections} considers the mathematical connections implied by (e).

  The DIDACKS approach was conceived and is best understood on its own merits as a self-contained mathematical theory that is independent of the geophysical connections just indicated and it is this exposition that occupies the first four sections.  Underpinning this mathematical side of the DIDACKS approach are three intrinsically interesting relationships that will now be briefly surveyed.  First consider the notation employed.   For overall accessibility, for consistency with the geophysical and electrostatic literature and to avoid various possible notational conflicts preference is given to a pedestrian but unambiguous notation:  Cartesian coordinates are used and overset arrows employed to denote \riii\ vectors, $\vec{X} = (x, y, z)^T \in  \mriii$ (for superscript $T \eq$ transpose), while for $n \neq 3$ $n$-dimensional vectors and matrices are denoted by lower and upper case bold letters, respectively.  $\text{R}_{\sps}$ is used to denote the radius of the sphere associated with $\Omega_{\sps}$ that is centered over the origin: $\Omega_{\sps} \eq \{\vec{X} \in  \mriii\mid |\vec{X}| \geq \text{R}_{\sps} \}$.  Likewise for the half-space case, the origin is chosen in the plane $\partial \Omega_1 \eq \{\vec{X} \in  \mriii\mid z = 0 \}$ so that $\Omega_1  \eq \{\vec{X} \in  \mriii \mid z \geq 0 \}$.  (Observe that the overall shape of the subscripts here match that of the associated boundary.)  Frequently these two settings will be denoted by a subscript $j$ (for example, $\Omega_j$), where $j = {\sps}$ or $1$ is always implied.  

   Temporarily leaving aside the issue of admissible functions, these relationships can be compactly stated in terms of a Dirichlet integral over some connected but possibly unbounded region $\Omega$, which are usually denoted by $\text{D}[v,\,w] = \iiint_\Omega \vec{\nabla} v\cdot\vec{\nabla} w \,\,dV$  for admissible harmonic functions $v(\vec{X})$ and $w(\vec{X})$, or in terms of the more inclusive concept of a weighted Dirichlet integral for the region $\Omega$ denoted by $\text{D}[v,\,w,\,\mu,\,\Omega]$, where $\mu = \mu(\vec{X})$ is the weighting function so that $\text{D}[v,\,w,\,\mu,\,\Omega] :=  \iiint_\Omega \mu\,\vec{\nabla} v\cdot\vec{\nabla} w \,\,dV$.  Clearly $\text{D}[v,\,w,\,1,\,\Omega] = \text{D}[v,\,w]$.  Let $\ell^{-1} \eq 1/|\vec{X} - \vec{X}'|$, where $\vec{X} \in \Omega_j$ and $\vec{X}'$ is in the corresponding closed source region $\eq \Omega_{S_j} \subset \Omega^{\prime}_j \eq$ compliment of ${\Omega}_j$. (By convention, generally primed variables occur in $\Omega^{\prime}_j$ and unprimed ones in $\Omega_j$.) Then the first two relations give the replication (or generalized collocation) properties of the DIDACKS kernel $\ell^{-1}$: 
\begin{equation}\label{E:mkeqn1}
 \text{D}[w,\,\ell^{-1},\,1,\,\Omega_1]\, =\, 2\pi\,w(x',\,y',-z')
\end{equation}
and
\begin{subequations}\label{E:mkeqn2}\\  
\begin{align}
 \text{D}[w,\,\ell^{-1},\,\mu_{\sps},\,\Omega_{\sps}]\, &=\, {2\pi\,|\vec{P}|\,\,w(\vec{P})}/{{\text{R}}_{\sps}^2}\ \,\,\tag{\ref{E:mkeqn2}a}\\
\intertext{with $\mu_{\sps} = 1/r$ ($\,r \eq |\vec{X}|$) and where}
\vec{P} &= \left(\frac {\text{R}_{\sps}^2}{|\vec{X}'|^2}\right) \vec{X}'.\tag{\ref{E:mkeqn2}b}
\end{align}
\end{subequations}
 Finally the third relationship ties the  unweighted Dirichlet integral over $\Omega_{\sps}$ to the weighted integral given on the left hand side (LHS) of (\ref{E:mkeqn2}a) and can be written as
\begin{subequations}\label{E:mkeqn3}\\  
\begin{align}
\text{D}[v,\,w,\,1,\,\Omega_{\sps}] &= {\text{R}_{\sps}}\cdot \text{D}[v,\,w,\,\mu_{\sps},\,\Omega_{\sps}] + \left(2\pi{\text{R}_{\sps}}\right)\cdot(v,\,w){\ls}_{\sigma}\ ,\tag{\ref{E:mkeqn3}a}\\
\intertext{where the surface inner product on the right hand side (RHS) here is defined as} 
(v,\,w){\ls}_{\sigma} &\eq (1/{4\pi})\,\iint\limits_{\sigma} v(r, \theta,\,\phi)\,w(r, \theta,\,\phi)\,d\,\sigma \tag{\ref{E:mkeqn3}b}
\end{align}
\end{subequations}
and where, as in \cite{HandM} and \cite{WandM}, $\sigma$ and $d\,\sigma$ 
have the following meaning when associated with the integral of $f(\vec{X})$
\begin{equation}\label{E:sigeqn}
 \iint\limits_{\sigma} f(r, \theta,\,\phi)\,d\,\sigma\, \eq \int\limits_{\phi=0\ \ }^{\ 2\pi}\negthickspace\negthickspace\!\!\int\limits_{\ \theta=0}^{\ \,\,\pi} \negmedspace\left[f(r,\,\theta,\,\phi)\right]{\Big|}_{r=R_{\sps}}\negmedspace\, \sin\,\theta\,\,d\,\theta\,d\,\phi\ 
\end{equation}
for standard spherical coordinates $r, \theta, \phi$. [Occasionally the limits implicit in (\ref{E:sigeqn}) will be stated explicitly for emphasis.]  Due to the way $r$ dependence enters in (\ref{E:sigeqn}) it can be used to derive expressions that are otherwise not obvious, as will be apparent in the sequel.

  Clearly (\ref{E:mkeqn1}) and (\ref{E:mkeqn2}a) give a means of performing closed-form inner products based on the Dirichlet integral, while (\ref{E:mkeqn3}a) links the weighted Dirichlet inner product to the unweighted one over $\Omega_{\sps}$.  The general approximation and functional analysis framework for these relationships is given in Section~\ref{S:GLLSQ}.  The derivation of (\ref{E:mkeqn1}) is given in Section~\ref{S:halfspace} and that of (\ref{E:mkeqn2}a) and (\ref{E:mkeqn3}a) in Section~\ref{S:sphere}.  In order to understand their connection to the applications described in Section~\ref{S:PMhist} and to put them in proper historical context, next briefly consider the history of these relationships.  Originally  (\ref{E:mkeqn2}a) and (\ref{E:mkeqn3}a) were discovered in a different form by the author in the early 1980's---namely (\ref{E:IntRep}) and (\ref{E:rel2}), where the ``integral norm'' introduced in Section~\ref{S:sphere} is used in place of the weighted Dirichlet integral.  The history of these relationships in this form and of their application from inception to the mid-1990's can be found in \cite{ruf}.  The germane part of this internal history is summarized and updated in Section~\ref{S:PMhist}.  As an aside, although (\ref{E:mkeqn1}) is not explicitly mentioned in \cite{ruf}, the derivation in Section~\ref{S:halfspace} of it is the one originally found by the author in the early 1980's.  The direct relationship of the integral norm to the weighted Dirichlet norm (\ref{E:inorm2d}) is new.  Also unless otherwise noted, the presentation itself (including various terms and concepts) is completely new here.  Finally, while the general DIDACKS technique is a synthesis of several  results that seem to have been overlooked by the broader scientific community, for any mathematical approach that directly touches on harmonic analysis, RKHS theory and gravimetric inverse source theory either precedents or specialized parallel lines of development would seem to be a necessity.  The known ones for \riii\ DIDACKS theory are addressed in Section~\ref{S:GCconnections}.  In one way or another all of the discussions of Section~\ref{S:GCconnections} pertain to the second DIDACKS relation in weighted Dirichlet form, (\ref{E:mkeqn2}a).  As discussed there, (\ref{E:mkeqn2}a) should clearly be viewed as going back to Krarup in \cite{Krarup}, since he derived it there in a directly equivalent form.  Aside from the author's work, there are no known instances of (\ref{E:mkeqn1}) and (\ref{E:mkeqn3}a), or for that matter for the second DIDACKS relation expressed in the integral norm form (\ref{E:IntRep}).

  To motivate a further exploration of the paper's scope and limits it is useful to envision the possible reactions of four typical classes of readers to the above relations.  First, consider an applied mathematician, physicist, engineer or other scientist who may be familiar with the material in the first four chapters of \cite{Jackson}, but is not yet a seasoned practitioner and may benefit by consulting \cite{Mate}, \cite{Moritz} or \cite{Kellogg}.   (The physical and historical importance of Dirichlet integrals and of the associated Dirichlet principle may be obtained from other sources \cite{Bergman,Mona,Garding}.)  This reader may find all three of the above field energy expressions somewhat surprising: (\ref{E:mkeqn1}) and (\ref{E:mkeqn2}a) because they allow for the closed form evaluation of volume integrals and (\ref{E:mkeqn3}a) since it allows for an unusual reexpression of the volume field energy.  This reader may also observe that (\ref{E:mkeqn1}) and (\ref{E:mkeqn2}a) appear to have some connection to Green's functions and the method of images \cite{Jackson} and may recognize (\ref{E:mkeqn2}b) as the coordinate transformation part of a Kelvin transformation \cite{Kellogg,Korns}.  These particular connections arise from the nature of Green's functions for $\Omega_j$, but the most direct explanation requires some knowledge of RKHS theory.  First, as noted in \cite{Bergman}, the existence of a closed-form reproducing kernel occurs when closed-form expressions for both Neumann and Dirichlet Green's functions exist.  Second, for the cases studied here a dual-action collocation kernel (DACK) arises from the result of a reflection, (\ref{E:mkeqn1}), or Kelvin transformation, (\ref{E:mkeqn2}b), applied to a reproducing kernel of the right form for the relevant geometry. 
    
  Second, while the possible reactions of any number of specialists might also be examined, consider a reader who has a particular interest in integral kernels or RKHS theory.  For various reasons this reader may also find the above relationships somewhat surprising.  This reader might, for example, observe that $\ell^{-1}$ plays the role of a kernel and then recall that a symmetric reproducing kernel (SRK) of the form $|\vec{X} - \vec{Y}|^{-1}$ for $\vec{X}$ and $\vec{Y}$ both in the same region cannot exist since a SRK must be bounded and this kernel is not.  Here $\vec{X}'$ is a fixed interior point and $\vec{X}$ is in the exterior region so $\ell^{-1}$ is bounded, which is a very different situation and implies a change of perspective.   This in itself could raise further questions since although DACKs are not reproducing kernels they have some properties analogous to them.  In the previous paragraph the kernels studied here were linked to reproducing kernels, but it is unclear whether minimum norm DACKs can arise in other ways.  Moreover, consideration of DACKs as a separate class of kernels also raises the question as to how they fit into our current overall understanding of kernel structures.  These and other issues of a general nature are outside the scope of the present article, but some specific topics that might interest the second type of reader, such as the exact definition of the replication and generalized collocation property mentioned above, are addressed in Section~\ref{S:GLLSQ}.

 Third, consider a reader who is very applications and results oriented.  Such a reader may be disappointed to discover that there is not a table containing numerical examples and results; however, a discussion of point mass and point dipole DIDACKS results and their associated applications settings can be found in Section~\ref{S:PMhist}, where global non-linear least squares (NLLSQ) results are emphasized.  Due to the variety and nature of regional gravity data, as well as other issues \cite{ruf}, no known easily replicated example provides generic benchmark results, which is normally an expectation for these tabulated examples.  Moreover, point source fitting problems are part of a general class of problems that are ``notoriously'' ill-conditioned and  problematic \cite[pp. 214-222]{Blakely} so that each problem encountered should be tackled on its own terms, which means that one or two simple table examples cannot serve to provide adequate implementation guidance.  Unfortunately, a thorough discussion of associated implementation strategies is outside the scope of the present article.  Also observe that a concrete example provides a replication check that can serve as a consistency test for implementors, but this point is largely superfluous here since the DIDACKS approach exhibits the generalized collocation property with respect to point sources and thus, when implemented correctly, any point mass or dipole fit replicates the point field data that was used to produce it in the first place to within allowed round-off error and thus any implementation serves as its own self-consistency test.    

 Fourth and finally, consider a reader whose primary interest is in the theory and application of Laplacian inverse source theory.  Since there are many shared implementation pitfalls common to both point source field modeling and inverse source estimation problems, the comments just made in the last paragraph are also relevant in this context.  While specific mathematical tools and implementation strategies are not discussed here, readers with solid applications experience should be able to make direct use of the formalism presented.  These readers may also be interested in the topic of continuous parameterized distributions, which is raised in Section~\ref{S:GLLSQ}.  Finally, it is worth noting that other source region shapes can be entertained within the contexts of the two considered geometries since the only real requirement is that source regions be bounded and contained within the compliment of the unbounded harmonic field region.

\section{Generalized Linear-Least Squares Setting}\label{S:GLLSQ}

This section addresses the generalized least-squares (GLLSQ) plan of approach and the associated functional space setting.  There are numerous approaches closely aligned to the GLLSQ one adopted here---such as the Galerkin and Raleigh-Ritz based techniques mentioned in the introduction; however, the acronym GLLSQ is introduced to imply an implicit change of perspective.  In particular, connections to generalized collocation, as discussed later in the section and explicitly formalized by the GLLSQ collocation condition, are implied as well as an approach that is distinct from the usual LLSQ ones where sampling and discretization are introduced.  Connections to GC are also implied.

 Both LLSQ and GLLSQ approaches minimize some cost function $\Phi' = \| v - w  \|^2$, where for the problems of interest $v(\vec{X})$ is a point source potential model and $w(\vec{X})$ is some given canonical (or truth) reference potential.  For a point mass fit with $N_k$ point masses $v$ has the form
\begin{equation}\label{E:pmpot}
 v(\vec{X}) = G\,\sum\limits_{k=1}^{N_k}\, \frac{m_k}{|\vec{X} - \vec{X}'_k|}\,\,,
\end{equation} 
where $G$ is the Newtonian gravitational constant $\approx 6.6742\times 10^{-11} \text{m}^3 \text{s}^{-2} \text{Kg}^{-1}$ \cite[p.3]{WandM}.   As previously noted, $\vec{X} \in \Omega_j$ and $\vec{X}'_k \in \Omega_{S_j}$, which is a bounded and closed subregion of the open region $\Omega'_j$, so that the (kernel) basis functions occurring in (\ref{E:pmpot}) are always bounded.  Further $\vec{X}'_{k'} \neq \vec{X}'_k$ for all $k' \neq k$ is always assumed. 

   Five conventions are adopted here.  First, in physics texts the potential function $v$ is interpreted as potential energy and (\ref{E:pmpot}) has a negative sign since all gravitational bodies attract and the resulting force is given by the negative of the gradient of the potential.  In geophysics these sign conventions are different and consistent with (\ref{E:pmpot}), but in either case this should cause little difficulty since in fitting problems all that is required is that the overall sign conventions for $v$ and $w$ be consistent.   Second, it is assumed that gravitational force is always acting on a unit test mass \cite[p. 4]{WandM} and thus it will be treated as having the units of acceleration \cite[p. 45]{WandM}.  Third, physical geodesists distinguish between gravity field quantities, which include the effects of the Earth's rotation, and gravitational quantities like (\ref{E:pmpot}), which do not \cite[p. 44]{WandM}.   This is a distinction physicists generally do not make since rotational effects can be easily tracked and accounted for as required.  The physicist's lead is followed here and this distinction is ignored.  Fourth, both positive and negative masses will be considered a possibility since this is the usual convention adopted in point mass fitting approaches.  Specifically, for gravity modeling and estimation problems each $m_k$ can clearly be viewed as a mathematical parameter that can assume either sign.  This convention also allows for the ready adaptation of material developed here to other areas where both signs can occur. (Even regional geoexploration inverse mass density estimation problems can be handled by assuming that all smoothed density estimates are with respect to an average or ambient density.)  Fifth, it is useful to introduce scaled versions of the above potential functions in order to absorb the factor of $G$: $ V = v/G$ and $ W = w/G$.  Thus the cost function to be minimized becomes (with $\Phi \eq \Phi'/G^2$):
\begin{equation}\label{E:cost}
\Phi = \| V - W  \|{\ls}^2 = \|V\|{\ls}^2 - 2\,(V,\,W) + \|W\|{\ls}^2\ . 
\end{equation}
(Notice that scaling a cost function leaves the minima unchanged.)

Next consider the philosophy behind the norm selection process.  As discussed later in Section~\ref{S:PMhist} the minimization philosophy of matching the observations as closely as possible has generally been chosen for point mass fitting problems.   This philosophy is, however, not necessarily sound in all or even most cases.  For modeling problems a reference model, which is assumed to be accurate, is given and one wishes to match this reference as closely as possible in some physical sense.  Here the desire is to minimize the possible error differences that will result when this given reference model is replaced by a new point mass (or point source) model, which invariably occurs in some sort of software emulation of a physical situation.  Thus instead of ``matching the observables as closely as possible,'' a sounder strategy is to ``minimize the type of errors that will lead to the greatest errors in the end product.''  From Newton's second law, since these end-product errors here are most often the direct result of gravity errors it is clear that the difference in the given gravity reference field and the developed point mass gravity model should be minimized, say in a squared residual sense at a large number of appropriate sample points.  As this distribution of sample points becomes uniformly dense over the entire global region of interest the following key integral condition results:  
\begin{equation}\label{E:Minimize}
 \text{Minimize}\ \  \Phi = \iiint\limits_{\Omega_j} |\vec{\nabla} V - \vec{\nabla} W|^2\, dV\, =\, \text{D}[W-V,\,W-V,\,1,\,\Omega_j]\ . 
\end{equation}

   Temporarily leaving aside fundamental issues, such as how to turn the RHS of (\ref{E:Minimize}) into a proper norm structure, consider the general form of the linear equation sets that result from minimizing this type of cost function.  For concreteness consider the minimization process in \riii\ half-space ($\Omega_{1}$).  Since the RHS of (\ref{E:Minimize}) is already proportional to the field energy, it is natural to consider the half-space ($j = 1$) energy norm:
\begin{equation}\label{E:genlsq}
\|V - W\|{\ls}_{{\text{E}}_1}^2 \eq\, \frac1{8\pi}\iiint\limits_{\Omega_1} |\vec{\nabla} V - \vec{\nabla} W|^2\,\,dV 
\end{equation}
(a factor of $8\pi$ has been inserted since it often occurs for various field energy expressions in appropriate units).
Because $\|V - W\|{\ls}_{{\text{E}}_1}^2 =\, \|V\|{\ls}_{{\text{E}}_1}^2 +\, \|W\|{\ls}_{{\text{E}}_1}^2 -\, 
 2(V,\,W){\ls}_{{\text{E}}_1}$, the energy inner-product
\begin{equation}\notag
 (V,\,W){\ls}_{{\text{E}}_1}\, \eq\,\text{D}[V,\,W,\,1,\,\Omega_1]/8\pi
\end{equation}
 is also needed.   In particular if $V$ is specified through (\ref{E:pmpot}), with ${\ell}_k \eq |\vec{X} - {\vec{X}}'_k|$, and if $W$ is an appropriate reference field, then 
\begin{equation}\label{E:llsqeqn}
\|V - W\|{\ls}^2_{{\text{E}}_1} =\, \|W\|{\ls}^2_{{\text{E}}_1} -\,  2\sum_{k=1}^{N_k} m_k({\ell}^{-1}_k,\,W){\ls}_{{\text{E}}_1} + \sum_{k=1}^{N_k}\sum_{k'=1}^{N_k} m_k\,m_{k'}({\ell}^{-1}_k,{\ell}^{-1}_{k'}){\ls}_{{\text{E}}_1}\ .
\end{equation}
Taking the partial of (\ref{E:llsqeqn}) with respect to $m_{k''}$ (for $k'' = 1,\,2,\,3,\,\ldots,\,N_k$), setting the result to zero, then dividing by two yields a linear equation set that can be easily inverted for the mass values, provided that $({\ell}^{-1}_k,\,\phi){\ls}_{{\text{E}}_1}$ can be easily computed for $\phi = W$ and $\phi = 1/{\ell}_k$.  The relationship which makes this possible is (\ref{E:mkeqn1}).  By introducing $T_{k, k'} = ({\ell}_k^{-1},{\ell}_{k'}^{-1}){\ls}_{{\text{E}}_1}$ and $A_k = (W,{\ell}_k^{-1}){\ls}_{{\text{E}}_1}$ the linear equation set can be written as
\begin{equation}\label{E:mkeqn}
 \sum\limits_{k'=1}^{N_k}T_{k, k'}\, m_{k'} = A_k\ .
\end{equation}

  For the spherical exterior matters proceed in much the same fashion except that a weight function, ${\mu}_{\sps} = 1/r$, must be introduced into (\ref{E:Minimize}).  Not only is this weighting required to turn ${\ell}_k^{-1}$ into a replicating kernel, but since regions closer to the Earth's surface are normally of greater interest for geophysical applications than regions further away it is also desirable.  

   Obviously with regards to applications (\ref{E:mkeqn}) is pivotal and, as such, warrants at least an informal examination.   Before proceeding it is useful to clarify the differences between reproducing, replication and generalized collocation kernels.  When a kernel, like ${\ell}_k^{-1}$, is not symmetric since its arguments are in different domains and there is a relationship such as (\ref{E:mkeqn1}) or (\ref{E:mkeqn2}a) that allows for closed-form inner-product expressions it will be called a replication kernel and be said to have the point replication property.  Since reproducing kernels are necessarily symmetric they cannot be considered replicating kernels so this terminology distinguishes replication and reproducing kernels.  Alternatively, the term generalized collocation property will be taken to be a generalization of a point data and/or collocation matching condition and as such includes not only the possibility of reproducing kernels, but also replicating kernels.  As discussed below, it also allows for the possibility that resulting inner products may be obtainable by numerical means (after assuming an underlying replication property also holds)---so long as the resulting inner products, $A_k$, occuring on the RHS of (\ref{E:mkeqn}) can be matched.   (The $A_k$'s may also represent empirically obtained data.) 

  As a concrete example of situations where numerical integration frequently enters consider fits based on the continuous analog of (\ref{E:pmpot}) where the potential is due to some parameterized density function $\rho$:
\begin{equation}\label{E:rhopot}
 V(\vec{X}) = \iiint\limits_{\Omega_{S}}\, \frac{\rho(\vec{X}',\,\mathbf{\alpha})}{|\vec{X} - \vec{X}'|}\,\,dV'\ .
\end{equation} 
Here $\alpha = ({\alpha}_1,\,{\alpha}_2,\,{\alpha}_3,\,\ldots,\,{\alpha}_{N_k})^T$.   If $\rho$ consists of a linear superposition of density basis functions ${\psi}_k$ \{i.e., $\rho = \sum_{k=1}^{N_k}{\alpha}_k{\psi}_k(\vec{X}')$\} then minimizing $\|W - V\|^2$ yields a linear equation set similar to (\ref{E:mkeqn}), where the $A_k$'s and $T_{k, k'}$'s must generally be computed numerically; however, when (\ref{E:mkeqn1}) or (\ref{E:mkeqn2}a) is used then great simplifications result and continuous distributions are tractable.  Besides parameterized volume distributions, parameterized surface and line distributions are also obviously possible.   One interesting choice for volume density basis functions ${\psi}_k$ is the use of finite element method (FEM) basis functions.  Thus consider the case where $\vec{q}^{\,\prime}_k$ are taken to be a set of node points over $\Omega_{S}$ and the ${\psi}_k(\vec{X}')$ are chosen to be a set of localized FEM basis functions with the property that ${\psi}_{k'}(\vec{q}^{\,\prime}_k) = 1$ if $k' = k$ and ${\psi}_{k'}(\vec{q}^{\,\prime}_k) = 0$ otherwise.  The ${\alpha}_k$ determined directly from the analog of (\ref{E:mkeqn}) then represent the density strengths at the node points ${\vec{q}}^{\,\prime}_k$.

   Any additional structure that can help to clarify the possibilities inherent in (\ref{E:mkeqn}) is desirable.  Towards that end briefly consider a suggestive theorem from RKHS theory.  When a reproducing kernel, with specified kernel points ${\vec{Q}}_k \in \Omega$ replacing the values of $\vec{X}'_{k}$ in (\ref{E:pmpot}), is used for a basis functions expansion of $V$ that is analogous to (\ref{E:pmpot}), and when (\ref{E:cost}) is replaced by the corresponding cost function based on the reproducing kernel norm, then minimization of this cost function results in a closed-form linear equation set that is just like (\ref{E:mkeqn})---except that the inner products corresponding to $A_k$ take on the simpler form $W({\vec{Q}}_k)$.  This type of reproducing kernel fit also satisfies a minimum norm collocation property:  the function with the smallest associated norm that matches the prescribed data set [i.e., the values $W({\vec{Q}}_k)$] is the one which results from solving the analog of (\ref{E:mkeqn})  \cite[pp. 207--220]{Moritz}.  This minimum norm property is a well known functional analysis result \cite{Mate} and it insures that a reproducing kernel fit will simultaneously match the given point data and minimize both $\| V\|$ and $\|V - W\|$ for the associated norm.  This fact is of interest here since it strongly suggests that if a replicating kernel expansion for $V$ is used in (\ref{E:mkeqn}), then generally any specified values for $A_k$ are recovered and that $\text{D}[V,\,V,\,1,\,\Omega_1]$ or $\text{D}[V,\,V,\,\mu_{\sps},\,\Omega_{\sps}]$ also is simultaneously minimized.  While this may fail to happen due to auxiliary restrictions placed on the basis functions or on the overall space of admissible functions, the main way that it can fail to happen is if the kernel basis functions themselves are not linearly independent.  These possibilities are not usually addressed in connection with general discussions of the minimum norm collocation property, but in many settings linear independence may not always be transparent, especially if combinations resulting from linear operators acting on kernel basis functions are allowed.  (These possibilities can be seen from, among other things, the consequences of the fact that various restricted classes of functions, such as polynomials of fixed degree, may have a reproducing kernel.)  When (\ref{E:mkeqn}) is invertible the source parameters ($m_k$) are uniquely determined and in some sense one can say that the solution to the inverse point source problem has been obtained.  To preserve and extend these inverse source interpretational possibilities, the conservative stance is adopted here of requiring that all admitted basis function sets be invertible and that the solutions to (\ref{E:mkeqn}) replicate the specified $A_k$ values.  This condition is called the generalized least squares collocation (GLLSQC) condition.  There are two obvious ways to enforce this condition: either on a computational case by case basis or by proving general theoretical results about classes of particular basis functions.  While it is not comparatively well known outside geophysics, a demonstration exists that shows point mass basis functions are independent in \R{n} for finite $n > 1$ \cite{geoPMuniq}.  Thus, in a theoretical sense the GLLSQC condition holds for point mass fits, but on a case-by-case basis some care is generally required in solving (\ref{E:mkeqn}) due to ill-conditioning.  (As a matter of practice, either a singular value decomposition or a Hausholder triangulation algorithm implemented to an appropriate number of significant digits should be used.)  Finally, it should be readily apparent that while continuous distributions may well satisfy the GLLSQC condition, counter-examples can be easily constructed, with perhaps the simplest example resulting from the consideration of several concentric homogenous spherical shells.

   With regard to the overall functional analysis setting, the standard course taken nowadays is to adopt some type of general Banach space setting (such as one type or another of Sobolev space), where the completeness of Cauchy sequences is presupposed.  The limit of sequences of functions composed from basis functions that satisfy the GLLSQC condition may not satisfy it, hence admitting limits of sequences can lead to unwanted interpretational difficulties here.  Obviously, to solve (\ref{E:mkeqn}) it is only necessary that a finite linear span of basis functions be admitted, which requires only an inner-product structure.  In accord with the conservative stance outlined above, a structured pre-Hilbert space setting is adopted since a pre-Hilbert space setting presupposes an inner product structure, but it does not make the usual assumptions about admissible sequences of functions.   (This way of specifying functions of interest is somewhat dated, but prior to the mid-1950's it was the prevailing way of addressing Dirichlet inner product spaces and it was used, for example, in \cite{Bergman}).  The adjective ``structured'' here means that further auxiliary conditions are imposed on the class of admissible functions.  One such condition is
that any set of basis functions considered must satisfy the GLLSQC condition stated above.  Another requirement is that all functions, $f$, must be harmonic over $\Omega_j$ (including $\partial\Omega_j$).  Additional structure is needed to insure that (weighted) Dirichlet integrals over unbounded domains can be regarded as defining a positive definite norm.  This can be accomplished by assuming one (or all) of the following four largely equivalent requirements:
\renewcommand{\theenumi}{\Roman{enumi}}  
\renewcommand{\labelenumi}{(\theenumi)} 
\begin{enumerate}
\item
$\text{D}[f,\,f,\,1,\,{\Omega}_j]$ is bounded and $f$ tails off at least as fast as $1/r$ as $r \longrightarrow \infty$.
\item
The first Sobolev norm of $f$ over ${\Omega}_j$ is bounded:  $\text{D}[f,\,f,\,1,\,{\Omega}_j] +  \iiint\limits_{{\Omega}_j} |f|^2\,d\,V < \infty$.
\item
$f$ is a potential function generated by a (well-behaved) localized source distribution in ${\Omega}_{S_j} \subset {\Omega}'_j\setminus\partial{\Omega}'_j $. 
\item
A well behaved series representation for $f$ always exists in the form of a spherical harmonic expansion of $f$ in powers of $1/r$, for the exterior of some sphere contained inside ${\Omega}'_j$.
\end{enumerate}
Notice that (I) is the historical approach \cite{Bergman} and that (II) equivalent to (I) for ${\Omega}_j = {\Omega}_{\sps}$ since both the integrals occurring in the first Sobolev norm must be separately bounded.  Also for all $f$ not identically zero, $0 < \text{D}[f,\,f,\,{\mu}_{\sps},\,{\Omega}_{\sps}] < \text{D}[f,\,f,\,1,\,{\Omega}_{\sps}]/R_{\sps}$, where positivity can be easily proved using several standard properties of harmonic functions. (Specifically, if $|f| \neq 0$ at some point in ${\Omega}_{\sps}$ then taking the line integral of $\vec{\nabla} f$ from this point to a point at infinity one can infer that $|\vec{\nabla} f| > 0$ for at least one point along this line. Then from the mean-value theorem and the maximum-modulus theorem $\vec{d}\cdot\vec{\nabla} f > 0$ must occur throughout some neighborhood near $\partial{\Omega}_{\sps}$ for one fixed direction $\vec{d}$ or another and from this it follows immediately that $\text{D}[f,\,f,\,{\mu}_{\sps},\,{\Omega}_{\sps}] > 0$.)  It is clear that (III) and (IV) are largely  equivalent to (I) or (II) [although (III) and (IV) do not require harmonicity as a separate condition].  Requirement (III) is more or less tantamount to the assumption that a collection of a point sources is being modeled.

\section{Half-Space ($\Omega_1$) Relationships}\label{S:halfspace}

For the class of admissible functions just described with bounded source region, the following half-space analog of Poisson's solution to the Dirichlet boundary value problem holds \cite[p. 268]{CourantII}:
\begin{equation}\label{E:posint}
W(\vec{X}) = \frac1{2\pi}\! \int_{\!-\infty}^{\,\,\infty} \int_{\!-\infty}^{\,\,\infty}
 \frac{W(x'', y'', 0)\,z\,\,dx''\,dy''}{[(x-x'')^2 + (y-y'')^2 + z^2]^{3/2}} 
\end{equation}
where $z  > 0$ and $\vec{X}'' \in \partial\Omega_1$.  Using the facts that the surface integral at infinity is zero for the unbounded region $\Omega_1$, that ${\partial\,}/{\partial\, n} \eq - \,{\partial\,}/{\partial z}$ and that $dS = \,dx\,dy$ for the boundary plane ($\partial\Omega_{1}$) and applying Green's first identity\footnote{\ $\iiint_{{\Omega}}(\phi{\nabla}^2\psi + \vec{\nabla}\psi\cdot\vec{\nabla}\phi)\,\,dV = \iint_{\partial\Omega} \phi\frac{\partial\psi}{\partial n}\,dS$} yields:
\newcommand{\infBOTlim}{{\text{\begin{small}{$\!-\infty$}\end{small}}}}
\newcommand{\infTOPlim}{{\text{\begin{small}{$\infty$}\end{small}}}}
\newcommand{\smallW}{\text{\begin{small}{$W$}\end{small}}}
\newcommand{\smallg}{\text{\begin{small}{$g$}\end{small}}}
\begin{equation}\label{E:dirform}
({\ell}_k^{-1},\,W){\ls}_{{\text{E}}_1} =\, \frac1{8\pi}\iiint\limits_{\Omega_1} \vec{\nabla} {\ell}_k^{-1}\cdot\vec{\nabla} W\,\,dV\, = 
 - \frac{1}{8\pi}\negthickspace{\text{\begin{large} ${\int_{\negmedspace -\infty}^{\,\infty} \int_{\negmedspace -\infty}^{\,\infty}} \left[ \smallW\,\tfrac{\partial \ }{\partial z}\tfrac1{|\vec{X} - {\vec{X}}'_k|}\right]{{\bigg|}_{z=0}}$\end{large}}}\!dx\,dy\ .
\end{equation}
Next observe that 
\begin{equation}\label{E:iden}
{\text{\begin{large} $\left[\tfrac{\partial \ }{\partial z}\tfrac1{|\vec{X} - {\vec{X}}'_k|}\right]{\biggr|}_{z=0}$\end{large} }}\negmedspace\! = \,\,\frac{z'_k\ }{[(x-x'_k)^2 + (y-y'_k)^2 + (z'_k)^2]^{3/2}}\ ,
\end{equation}
where $z'_k < 0$. Combining (\ref{E:posint}), (\ref{E:dirform}) and (\ref{E:iden}) produces:
\begin{equation}\label{E:e2rk}
({\ell}_k^{-1},\,W){\ls}_{{\text{E}}_1} = {W(x'_k,\,y'_k,\,-z'_k)}/4\ ,
\end{equation}
which shows (\ref{E:mkeqn1}).  This relation can be used to evaluate the terms $T_{k, k'}$ and $A_k$ occurring in (\ref{E:mkeqn}):
\begin{equation}\label{E:E1TKAK}
  T_{k, k'}\, = \,\frac1{4\,\sqrt{(x'_k - x'_{k'})^2 + (y'_k - y'_{k'})^2 + (z'_k + z'_{k'})^2}}\,,\ \ \ A_k = \frac{W(x'_k,\,y'_k,\,|z'_k|)}4\ .
\end{equation}

  Two final points are relevant.  First, dipole and other higher order multipoles also can be easily fit.  In general potentials for these point sources can be written as $\sum_k\sum_i S_{ki}\mathscr{L}_{ki}({\ell_k}^{-1})$ where $S_{ki}$ are the associated source strengths and the $\mathscr{L}_{ki}$ are appropriate linear differential operators that can be expressed as a sum of partials of various orders with respect to $x$, $y$, or $z$.  For example, an electrostatic or point mass dipole term is proportional to $\vec{D}_k\cdot\vec{\nabla}{\ell^{-1}_k}$.  Since $\vec{\nabla}{\ell^{-1}_k} = - \vec{\nabla}'_k{\ell^{-1}_k}$, where $\vec{\nabla}'_k \eq ({\partial\,}/{\partial {x'_k}},\,{\partial\,}/{\partial {y'_k}},\,{\partial\,}/{\partial {z'_k}})^T$, and components of $\vec{X}'_k$ serve only as parameters when they occur inside the inner product here, all $\vec{X}$ dependent derivative factors operating on ${\ell^{-1}_k}$ that occur inside inner products can be replaced with $\vec{X}'_k$ derivative factors; hence, these differential operators can be moved inside or outside the inner products entirely as desired so that all required inner products for $T_{k, k'}$ and $A_k$ can be easily evaluated in closed form.   For half-space, these possibilities and the associated measurable point quantities are displayed in Table~\ref{Ta:MeasSour}.  Analogous possibilities exist for the spherical exterior case.

\begin{table} 
\begin{center}
\begin{tabular}{|c|c|c|}\hline
    Observation Type & Source Type      &  Invariance Property   \\ \hline\hline
    Potential        & Point Mass       &  Scalar                \\ \hline
    Gravity          & Point Dipole     &  Vector                \\ \hline
    Gravity Gradient & Point Quadrupole &  2nd Rank Tensor       \\ \hline
\end{tabular}
\caption{Point Data/Source Correspondences for \riii\ Half-Space}\label{Ta:MeasSour}
\end{center}
\end{table}

  Second, to improve the condition number of the matrix $\boldsymbol{T}$ in (\ref{E:mkeqn}), it is often desirable to use normalized basis functions $\hat{\varphi}_k$ in place of $\ell^{-1}_k$:
\begin{equation}\label{E:ubasis}
\hat{\varphi}_k \eq \frac{\widetilde{N}_k}{\ell_k}\,,\ \ \text{where}\ \ \widetilde{N}_k \eq \frac{1}{\|{\ell}^{-1}_k\|}
\end{equation}
in the general norm setting.  Introducing $\widetilde{T}_{k, k'} \eq (\hat{\varphi}_k,\,\hat{\varphi}_{k'}) = \widetilde{N}_k\,\widetilde{N}_{k'}\,T_{k, k'}$, $\tilde{A}_k \eq (W,\,\hat{\varphi}_k) = \widetilde{N}_k\,A_k$ and $\widetilde{m}_k \eq m_k/\widetilde{N}_k$ allows (\ref{E:E1TKAK}) to be reexpressed as
\begin{equation}\label{E:unitGLLSQ} 
\sum\limits_{k'=1}^{N_k}\widetilde{T}_{k, k'}\, \widetilde{m}_{k'} = \tilde{A}_k\ .
\end{equation}
For the \riii\ half-space energy norm 
\begin{equation}\label{E:unitTMA}
\widetilde{N}_k = 2^{\frac32}\sqrt{|z'_k}|\,,\ \ \widetilde{T}_{k, k'} =\frac{2\sqrt{z'_kz'_{k'}}}{\sqrt{(x'_k - x'_{k'})^2 + (y'_k - y'_{k'})^2 + (z'_k + z'_{k'})^2}}\,,\ \ \tilde{A}_k = \frac{\sqrt{|z'_k|}}{\sqrt{2}}\,W(x'_k\,\,y'_k,\,-z'_k)\,,
\end{equation}
so that (\ref{E:unitGLLSQ}) can easily be inverted to determine the values of $\widetilde{m}_k$ and thus $m_k$.  While the use of normalized basis functions is not always required for point mass fits, their use in mixed type point source fits is always highly recommended due to the diverse associated physical scales that occur there and the attendant large condition numbers for $\boldsymbol{T}$.  (An experiment was performed on the results from one of the global NLLSQ combined point mass/dipole fits discussed in Section~\ref{S:PMhist}.  At the specified source locations a linear fit was done both with and without normalized basis functions.  The ratio of the two resulting condition numbers was over $10^{20}$.)

\section{Spherical Exterior ($\Omega_{\sps}$) Relationships}\label{S:sphere}

The goal of this section is the derivation of the two relevant spherical exterior DIDACKS relationships, (\ref{E:mkeqn2}a) and (\ref{E:mkeqn3}a).  Matters are more complex for this case than they were for the half-space case, but all of the supporting issues raised in Sections~\ref{S:GLLSQ} and \ref{S:halfspace} can obviously be carried over here, so they are not repeated.

   Let $f$ and $g$ be two admissible functions as discussed in  Section~\ref{S:GLLSQ} and consider the integral norm \cite{ruf}: 
\begin{equation}\label{E:intnorm}
(f,\,g){\ls}_{\text{I}} \eq \,-\,\frac{R_{\sps}^2}{4\pi}\negthinspace\iint\limits_{\sigma} \Dr (rf\,g)\,\, d\,\sigma  = -\,\frac{R_{\sps}^2}{4\pi}\negthinspace\iint\limits_{\sigma} \big[ \Dr (f\,r\,g)\big]{\Big|}_{r=R_{\sps}}\negmedspace d\,\sigma
\ ,\ \text{ where } \Dr \eq \frac{\partial\ }{\partial r}\ \cdot
\end{equation}
The last expression on the RHS here follows from the evaluation convention of (\ref{E:sigeqn}).  The label ``integral norm'' was chosen by the author since, as discussed below, the integrals required for point mass fitting in $\Omega_{\sps}$ can be evaluated in closed form and there is little chance of confusing this norm with the usual norm of square integrable functions.   
 
 Applying Green's first identity to the $\Omega_{\sps}$ energy inner product and noting that the surface integral at infinity vanishes, while $dS = {R_{\sps}^2}\,d\,\sigma$ and ${\partial\,}/{\partial\, n} \eq -\Dr$ for the bounding inner exterior surface of $\Omega_{\sps}$, yields:
\begin{equation}\label{E:E0norm}
 (f,\,g){\ls}_{\text{E}_{\spss}} \eq \,\frac1{8\pi}\negthinspace\iiint\limits_{\Omega_\sps} \vec{\nabla} f\cdot\vec{\nabla} g\,\,dV =
-\frac{R_{\sps}^2}{8\pi}\negthinspace\iint\limits_{\sigma} g\,\Dr\!f\,\,d\,\sigma 
= -\frac{R_{\sps}^2}{16\pi}\negthinspace\iint\limits_{\sigma} \big[\Dr (fg)\big]{\Big|}_{r=R_{\sps}} d\,\sigma \ . 
\end{equation}
From (\ref{E:intnorm}) and (\ref{E:E0norm}) it follows immediately that
\begin{equation}\label{E:rel2}
 (f,\,g){\ls}_{\text{I}} = 4\,R_{\sps}(f,\,g){\ls}_{\text{E}_{\spss}} - \,R_{\sps}^2(f,\,g){\ls}_{\sigma}\ .
\end{equation}
After the second DIDACKS relation, (\ref{E:mkeqn2}a), is addressed, it will be shown that the integral norm is proportional to the weighted Dirichlet integral, which will complete the proof of the third DIDACKS relationship, (\ref{E:mkeqn3}a).

    First, observe that the following two equations can be shown through a relatively straightforward evaluation of their respective LHS and RHSs:
\newcommand{\oneoverXP}{\mLarge{\tfrac1{|\vec{X} - \,{\vec{P}}_k|}}}
\begin{equation}\label{E:Aidintnormen}
{\Blbrac}\frac{r}{{\ell}_k}{\Brbrac}\Biggr|_{r=R_{\sps}}\! = \
\frac{R_{\sps}^2}{r'_k}{\Blbrac}\!\oneoverXP\!{\Brbrac}{\Biggr|}_{r=R_{\sps}}\ 
\end{equation}
and
\begin{equation}\label{E:Biden}
{\Blbrac}\Dr\!\left(\frac{r}{{\ell}_k}\right){\Brbrac}\Biggr|_{r=R_{\sps}}\! = \
-\frac{R_{\sps}^2}{r'_k}\Blbrac\Dr\!\oneoverXP\Brbrac{\Biggr|}_{r=R_{\sps}}\ 
\end{equation}
where ${\vec{P}}_k$ is given by (\ref{E:mkeqn2}b) with ${\vec{X}}'_k = {\vec{X}}'$ and $r'_k \eq |{\vec{X}}'_k|$.  Employing (\ref{E:Aidintnormen}) and (\ref{E:Biden}) yields
\begin{equation}\label{E:Diden}
-\frac{R_{\sps}^2}{4\pi}\negthickspace\negmedspace\mLarge{\iint\limits_{\sigma}}\negthickspace\Blbrac \Dr\!\left(\frac{rW}{{\ell}_k}\right)\Brbrac\Biggr|_{r=R_{\sps}}\!\!d\sigma \ = \ -\frac{R_{\sps}^4}{4\pi r'_k} \negthickspace\negmedspace\mLarge{\iint\limits_{\sigma}}\negthickspace
\Blbrac {\mLarge{\tfrac{(\Dr W)}{|\vec{X} - \,{\vec{P}}_k|}}} - \,\,W\,\Dr\!\oneoverXP\Brbrac {\Biggr|}_{r=R_{\sps}}\!\! d\sigma\ .
\end{equation}
Recall that $W$ is harmonic for $r > R_{\sps}$.
Applying Green's second identity\footnote{\ $\iint_{\partial{\Omega}} \left(\phi\frac{\partial\psi}{\partial n} - \psi\frac{\partial\phi}{\partial n}\right) \,dS = \iiint_{{\Omega}}(\phi{\nabla}^2\psi - \psi{\nabla}^2\phi)\,\,dV$} to (\ref{E:Diden}) yields
$$
 -\frac{R_{\sps}^2}{4\pi}\negthickspace\negmedspace\mLarge{\iint\limits_{\sigma}}\negthickspace\Blbrac \Dr\!\left(\frac{rW}{{\ell}_k}\right)\Brbrac\Biggr|_{r=R_{\sps}}\!\!d\sigma \ =  - \frac{R_{\sps}^2 }{4\pi r'_k }\negmedspace\mLarge{\iiint\limits_{\Omega_{\sps}} }\negmedspace\! W\,{\nabla}^2\!\left(\negmedspace\!\oneoverXP\negmedspace\!\right) \,dV\ . 
$$
Then using ${\nabla}^2\left({1}/{|\vec{X} - {\vec{P}}_k|}\right) = -4\pi \delta (\vec{X} - {\vec{P}}_k)$, where $\delta$ is the Dirac delta function \cite[p. 35]{Jackson}, gives
\begin{equation}\label{E:Irep}
-\frac{R_{\sps}^2}{4\pi}\negthickspace\negmedspace\mLarge{\iint\limits_{\sigma}}\negthickspace\Blbrac \Dr\!\left(\frac{rW}{{\ell}_k}\right)\Brbrac\Biggr|_{r=R_{\sps}}\!\!d\sigma \ = \, \frac{R_{\sps}^2 }{r'_k }  W\left({\vec{P}_k}\right)  
\end{equation}
or finally, with $P_k = |{\vec{P}_k}|$:
\begin{equation}\label{E:IntRep}
({\ell}_k^{-1},\,W){\ls}_{\text{I}} \, =\, {P_k }\,  W\left({\vec{P}_k}\right)\ . 
\end{equation}

Various other ways to prove (\ref{E:IntRep}) exist.  One way is to substitute spherical harmonic expansions for $W$ and ${\ell}_k^{-1}$ into the LHS of (\ref{E:Irep}).  A second way is to first observe that $({\ell}_k^{-1},\,{\ell}_{k'}^{-1}){\ls}_{\text{I}}$ can be evaluated in closed form by using spherical coordinates on the LHS of (\ref{E:Irep}).  Next, this result can be generalized by substituting the integral form of Poisson's equation (not Poisson's integral) for $W$ into the LHS of (\ref{E:IntRep}) and then reintroducing the closed-form expression just obtained for $({\ell}_k^{-1},\,{\ell}_{k'}^{-1}){\ls}_{\text{I}}$\,, which results in the RHS of (\ref{E:IntRep}) reexpressed in terms of the integral form of Poisson's equation.  As previously noted these various steps were the ones originally followed by the author and account for the nomenclature.  Other proofs have also been discovered.

  This leaves the proof of the relationship between the weighted Dirichlet integral and the integral norm.  Recalling the limit convention implicit for integration over $\sigma$, (\ref{E:sigeqn}), and expanding the RHS of (\ref{E:intnorm}) yields 
\begin{equation}\label{E:inorm2}
 (f,\,g){\ls}_{\text{I}} =  - \frac{R_{\sps}^2}{4\pi}\negthinspace\iint\limits_{\sigma} f\,g\,\, d\,\sigma - \frac{R_{\sps}^2}{4\pi}\negthinspace\iint\limits_{\sigma} r\Dr (f\,g)\,\, d\,\sigma \ .
\end{equation}
Next temporarily ignore the common factor of ${R_{\sps}^2}/{4\pi}$ and consider the two integrals on the RHS of (\ref{E:inorm2}).  Using the fact that $\Dr f = ({\vec{X}}/{r})\!\cdot\!\vec{\nabla}f$, the first integral on the RHS of (\ref{E:inorm2}) can be written as
\begin{equation}\label{E:inorm2a}
\begin{split}
 - \iint\limits_{\sigma} f\,g\, d\,\sigma &= \iint\limits_{\sigma} \left\{\,\, \int\limits_{\,r=R_{\sps}}^{\ \ \infty}\negmedspace\!\Dr\,(f\,g)\,dr\,\right\}\, d\,\sigma\,\, =  \iiint\limits_{\Omega_{\sps}} r^{-2}\,\Dr\,(f\,g)\,\,dV
\\
 &=  \iiint\limits_{\Omega_{\sps}} (g\,\,r^{-3})\,({\vec{X}}\!\cdot\!\vec{\nabla}f)\,\,dV\,  + \iiint\limits_{\Omega_{\sps}} (f\,\,r^{-3})\,({\vec{X}}\!\cdot\!\vec{\nabla}g)\,\,dV \ .\\
\end{split}
\end{equation}

  Green's second identity in the following form will be useful in reexpressing the second term on the RHS of (\ref{E:inorm2}):
\begin{equation}\label{E:green}
\iiint\limits_{{\Omega}_{\sps}} \vec{\nabla}\psi\!\cdot\!\vec{\nabla}\phi\,\,dV = - \iint\limits_{\sigma} r^2\,\psi\,(\Dr\phi)\, d\sigma 
\end{equation}
where $\phi$ is harmonic in $\Omega_{\sps}$, but $\psi$ need not be and both must vanish sufficiently fast as $r \longrightarrow \infty$.
Using this identity at the appropriate place twice in the following expression (first with $\psi = f/r$ and $\phi = g$ and then with $f$ and $g$ reversed), the second term on the RHS of (\ref{E:inorm2}) can be rewritten as
\begin{equation}\label{E:inorm2c}
 - \iint\limits_{\sigma} r f\,(\Dr g)\, d\,\sigma - \iint\limits_{\sigma} r\,g\,(\Dr f)\, d\,\sigma = \iiint\limits_{\Omega_{\sps}} \vec{\nabla}g\!\cdot\!\vec{\nabla}\left({f}/{r}\right)\,\,dV
  + \iiint\limits_{\Omega_{\sps}} \vec{\nabla}f\!\cdot\!\vec{\nabla}\left({g}/{r}\right)\,\,dV\ . 
\end{equation}
Finally using $\vec{\nabla}(f/r) = r^{-1}\vec{\nabla}f - r^{-3}\vec{X}f$ in (\ref{E:inorm2c}) and substituting this result along with (\ref{E:inorm2a}) into (\ref{E:inorm2}) produces
\begin{equation}\label{E:inorm2d}
(f,\,g){\ls}_{\text{I}} =
\frac{R_{\sps}^2}{2\pi}\negthinspace\iiint\limits_{\Omega_{\sps}} r^{-1}\, \vec{\nabla}f\!\cdot\!\vec{\nabla}g\,\,dV\ . 
\end{equation}
With the aid of (\ref{E:rel2}), (\ref{E:IntRep}) and (\ref{E:inorm2d}), the second and third DIDACKS relationship [i.e., (\ref{E:mkeqn2}a) and (\ref{E:mkeqn3}a)] follow immediately.  From (\ref{E:inorm2d}) and the discussion at the end of Section~\ref{S:GLLSQ} the integral norm is positive definite.

  DIDACKS dipole and other higher order multipole implementations differ for the half-space and spherical settings in one significant way.  While closed-form expressions for all the required inner products for an integral-norm point-mass fit can be evaluated just as discussed in Section~\ref{S:halfspace}, for higher-order multipole fits all the derivatives of the potential for all the lower orders are also required in the spherical case because taking partials with respect to the components of ${\vec{X}}'_k$ yields additional terms on the RHS of (\ref{E:IntRep}).  This means, for example, that a dipole fit requires not only point gravity information, but point potential information as well.  It is thus natural in this case to perform not only a point dipole fit, but a combined point mass/dipole fit.  With this understanding the spherical case analog of Table~\ref{Ta:MeasSour} is readily obtained.

  When an integral norm higher order multipole fit is desired and the lower order derivatives of $W$ are not available then it still may be possible to do the fit \cite{ruf}.
For example, if a dipole fit is desired and information about $W$ is not available, but the gradient of $W$ is known along various intersecting lines, then potential data in the form $W(\vec{X}) = W_0 + \delta W(\vec{X})$, where $W_0$ is an unknown constant, can be assumed anywhere along these lines since the form of $\delta W(\vec{X})$ can be found by numerical line integration.  For any assumed trial value of $W_0$, a DIDACKS fit can be performed.  The results of this fit can then, in turn, be substituted into a standard LLSQ type of cost function that is the analog of (\ref{E:anomeqn}) and is based on minimizing gravity computations at various sample points.   If gravity data itself is plentiful, then an outer-loop optimization process can be based on minimizing this new cost function, where $W_0$ is treated as an unknown NLLSQ parameter.  In this outer-loop process sample point gravity differences are minimized throughout the fit region of interest (which may be only a small part of ${\Omega}_{\sps}$).  When its choice is not obvious, $R_{\sps}$ can also be treated as a parameter and optimized in the same fashion.  

\section{Broader Point Mass Fitting Context}\label{S:PMhist}

   The goal of this section is to illuminate certain aspects of relevant geophysics and point mass fitting background material by way of a brief synopsis (as such this section is somewhat subjective).   Before considering the associated geophysical context it is useful to clarify the distinction between NLLSQ and GLLSQ or LLSQ problems and to briefly consider how the pertinent DIDACKS applications history fits into these categories.  In the sequel, the distinction between GLLSQ and LLSQ is generally dropped and only the acronym LLSQ is used so that both classes can be referred to jointly.  If all the source locations are known then linear equation sets result for the source strengths and in the DIDACKS approach these linear equation sets are exact due to (\ref{E:mkeqn1}) and (\ref{E:mkeqn2}a).  Alternatively, if both the locations and strengths are to be determined then a NLLSQ problem results since source locations enter as non-quadratic parameters in the cost function.  While only LLSQ problems have been addressed so far in this article, accurate low degree and order spherical harmonic (tesseral) NLLSQ fits have been obtained by the author and this was, in fact, the first area of DIDACKS applications in the early 1980's \cite{ruf}.  These NLLSQ fits are discussed further below.  While rather varied approaches have been used by different researchers for point mass based geoexploration inverse source applications, for the gravity modeling and estimation problems dealt with here far fewer point mass based approaches have been employed; nevertheless, not only have the associated research efforts been internationally diverse, but the corresponding literature is also extensive so that only a small part of it can be considered here (see \cite{Vermeer} for additional history and references).  LLSQ and NLLSQ point mass gravity models can also be produced to serve as synthetic gravity models with realistic attributes, but DIDACKS applications in this area have not been considered  so it is not discussed below.  Finally, two conventions are adopted in the sequel: (a) a spherical harmonic expansion to degree and order $N$ will be called a $N \times N$ field or expansion and (b) both field modeling and estimation problems will be referred to as field reconstruction problems.

  Next, consider the relevant geophysical aspects.  Approximately a quarter of a century ago one could divide the Earth's gravity field into three parts corresponding to their respective data sources:
\begin{enumerate}
\item
 Global spherical harmonic field data derived directly from satellite tracking data and historically considered accurate to around the degree and order $8$ to $12$ range.  Here this part of the field is called `low degree and order'' and it is taken to be $12 \times 12$ and below.  
\item
An intermediate field taken here to be the part above $12 \times 12$ and below $120 \times 120$, which before the evolution of more advanced radar equipped satellites could not be accurately determined. 
\item
 Regional measurements of geophysical surface quantities known as gravity anomaly and vertical deflections. [These data sets contain part (I) and (II) contributions unless they are factored out.] 
\end{enumerate}
Parts (I) and (II) together will be taken as comprising the global part of the gravity field.  As discussed below, much higher accuracy is desirable for part (I) than part (II) or part (III) data.  Currently very accurate spherical harmonic expansions to a much higher degree and order are available on the Internet, so the distinction between the above three data sets is not as distinct as it once was, but to understand the context of the methods discussed in this section it is useful to keep this historical data partition in mind.  \{Expansions beyond $360 \times 360$ are common and the recent Gravity Recovery and Climate Experiment (GRACE) \cite{WandM,GRACE} has already established new global gravity accuracy benchmarks up through $110 \times 110$.\}

  Next consider the origins of part (III) data.  Regional data measurements are often made at sea and it is in this context that the concepts of gravity anomaly and vertical deflection components are best understood.  Suppose that the Earth were composed of a homogenous liquid with the same overall mass and volume, then due to rotational effects it would take on an ellipsoidal shape.  Geophysicists set up a mathematical model of this configuration that they call the reference ellipsoid. The associated field is called the normal gravity field and it is the predominant part of the field.   If transient effects (like tides and ocean waves) are accounted for, then under the influence of gravity the oceans form an equal-potential surface (otherwise water would flow from one part to another until an equilibrium was reestablished).  If normal gravity is subtracted off and if these transient effects are properly taken into account then the measurement taken by a gravimeter on a ship is called gravity anomaly, $\Delta g$, and it is a scalar since the measurement is taken along a plumb line.  The angular displacement of this plumb line from the vertical is called vertical deflection, which can be resolved into north-south and east-west components.  Due to the equipotential effect just noted, this measurement is displaced from the specified reference ellipsoid by a distance that is called geoid height, but the measurement is recorded as if it had been made at a point on the reference ellipsoid itself \cite{WandM}.  Geoid height and scalar potential values are connected by Bruns formula \cite{WandM}.  Geoid height itself can be determined by a radar equipped satellite that has a known location.  Gravity anomaly is typically measured in units of milligal, where $1$ milligal $= 1 \times 10^{-5} m/s^2$.  A $1$ milligal error is considered more-or-less acceptable for regional gravity anomaly data processing \cite[p. 274]{WandM} and various underpinning geophysical relationships are derived with an inherent approximation consistent with this $1$ milligal requirement \cite{HandM,WandM}.  Much higher accuracy is desirable for part (I) than part (II) or part (III) data since gravitational effects are cumulative for most uses and low degree and order errors tend not to cancel out.  GRACE and other modern fields have error levels considerably under a milligal for part (I) data and for this part of the field it is also desirable to have point mass models with errors somewhat under a milligal.

   Historically, there have been three primary motivations for performing NLLSQ fits to the low degree and order spherical harmonic part of the Earth's gravity field:  (a)  To find a more efficient computational scheme for gravity evaluations.  (b)  To gain some insight into the distribution of matter in the Earth's interior.  (c)  To conduct goal oriented pure research.  Due to the computational ease and speed of low degree and order spherical harmonic gravity evaluations that has resulted from computer hardware and software advances, (a) has long ceased to be a realistic reason for using point mass fits and this point is totally irrelevant now.  Other measurement programs and advances have also emerged to address the issues raised by (b), but in fact it was never clear that the small number of point masses used in NLLSQ fits could provide truly significant mantle or deep core density information, which leaves only (c).  NLLSQ point mass fitting presents a very challenging problem that can conceivably serve as a sort of test bed for developing techniques to tackle other ill-conditioned problematic NLLSQ problems; moreover, aside from these NLLSQ aspects, it is an intrinsically interesting potential theory problem that has associated cross-fertilization possibilities.  

  The first step in attacking any LLSQ or NLLSQ problem is to set up a cost function.  A commonly chosen minimization philosophy for the NLLSQ point mass fitting problem is that of matching the observed quantities as closely as possible.  Since there is a classical result in geophysics (Stokes' integral) that says that if gravity anomaly is known over the entire reference ellipsoid then the  external field quantities can be reconstructed, standard approaches to performing global NLLSQ low degree and order point mass fits often have been based on matching gravity anomaly at $N_i$ different sample points, $\vec{X_i}$, specified on the reference surface.  This technique will be called the ``classical point mass'' approach.  If $\Delta g_{PM}(\vec{X_i})$ denotes the anomaly generated by a collection of $N_k$ point masses and ${\Delta g_{ref}}(\vec{X_i})$ represents the truth anomaly value at the same point, this classical NLLSQ point mass approach can be framed through the requirement that the following cost function be minimized: 
\begin{equation}\label{E:anomeqn} 
\Phi = \sum_{i=1}^{N_i}\boldsymbol{\big(}\,\Delta g_{PM}(\vec{X_i}) - \Delta g_{ref}(\vec{X_i})\,\boldsymbol{\big)}^2\,, 
\end{equation}
where $N_i  >> N_k$.  The usual prescribed number of point masses, $N_k$, is approximately $50$ for a NLLSQ $9 \times 9$ fit, while $N_k \approx 80$ for a $12 \times 12$ fit.  The resulting gravity anomaly error standard deviations (sigma) achieved by applying this classical point mass fitting approach has typically been from $3/4$ to several milligals.  These approaches normally have a much more sizeable error in the smaller degree and order terms than is desirable.  

 The global NLLSQ point mass fitting problem itself is inherently nonlinear and the point masses must be quite deep to obtain good results, which heightens numerical difficulties and associated convergence problems.  Given this state of affairs, coupled with the facts that the smaller degree and order errors cannot be easily removed using this classical approach and that it contains inherent sampling and discretization error, there is an error floor of around $2/3$ milligal that these approaches historically have not been able to overcome.  While the associated regional LLSQ approaches that have been tried are remarkably diverse, far fewer low degree and order NLLSQ approaches have been employed---nevertheless the diversity of attempted NLLSQ point mass approaches in the literature is much wider than the above discussion indicates, but still the accuracy levels and deficiencies of the classical point mass approach presented above are thought to be representative of these other existing NLLSQ attempts as well.  In Section~\ref{S:GLLSQ} it was argued that the usual philosophy of matching  measured quantities is not the correct one and that an energy basis or weighted energy basis is clearly called for in approaching both LLSQ and NLLSQ gravity modeling problems.  

   As noted above, accurate global NLLSQ low-degree and order spherical-harmonic point-mass fits using the DIDACKS approach were first obtained almost a quarter of a century ago by the author and in the past as  better spherical harmonic data has became available more accurate fits have been obtained \cite{ruf}.  This has been an ongoing effort and the current NLLSQ $50$ point mass fit to the $9 \times 9$ part of a recent field has a sigma error of about $0.035$ milligal, while the corresponding error in an NLLSQ $80$ point mass fit to the $12 \times 12$ part of the field is $0.030$ milligal.  As one might expect, additional masses here can offer marked improvements in accuracy, but at a loss of efficiency.  Combined point mass/dipole fits (c.f., Section~\ref{S:sphere}), which are combinations of point masses and dipoles at the same location, have also been performed over the years \cite{ruf}.  Only about half as many of these combined point sources are required for an accurate NLLSQ fit: $22$ for a $9 \times 9$ field and $35$ for a $12 \times 12$ field.  These NLLSQ DIDACKS fits have all been based on the integral norm introduced in Section~\ref{S:sphere}.  Besides the basic DIDACKS formalism, these fits also use additional specialized NLLSQ techniques developed by the author to handle the existence of numerous false minima at various physical scales.  

   While the fits should be doable, neither LLSQ nor NLLSQ DIDACKS fits to the complete global intermediate part of the field [part (II)] have been attempted, although efficient and accurate DIDACKS fits for various regional and local gravity modeling and estimation problems using both LLSQ and partial NLLSQ implementations have been obtained by the author \cite{ruf}.  Regional point mass gravity field reconstruction [part (III)] based on the ``classical point mass'' approach epitomized by (\ref{E:anomeqn}) has also often been successfully done over the years by various researchers; however, some have reported disappointing results.  This is not surprising since considerable patience is also often required in order to obtain the best or even good results with the DIDACKS approach due to source placement issues.  Thus while both DIDACKS and the classical point mass fitting approach can be viewed as an alternative to GC for regional applications, they both clearly share the same sensitivity to and dependence on point mass placement.  This sensitivity to point mass placement is also apparent from the low degree and order NLLSQ DIDACKS results since NLLSQ iterations have obviously reduced the errors by several orders of magnitude over that of initial trial configurations.  GC is designed to not be as sensitive to the placement of the kernel measurement points (which is the corresponding placement issue for it) and thus GC generally requires less patience and skill.  Alternative part (II) and (III) grid-based point mass approaches are referenced and discussed in \cite{Vermeer}.  
 
   It is thus useful to briefly describe GC in order to compare and contrast it with the DIDACKS approach.  Unlike DIDACKS, GC theory generally assumes all available data is used.  GC as commonly practiced differs from standard collocation techniques utilized by applied mathematicians in five basic ways: 
\renewcommand{\theenumi}{\arabic{enumi}}  
\renewcommand{\labelenumi}{\theenumi} 
\begin{enumerate}
\item 
 A SRK basis is assumed and emphasis is primarily focused on a statistical (covariance) interpretation given to these kernels.
\item 
 Laplace's equation in \riii\ is always assumed to hold over the region of interest.
\item 
 This field region is assumed to be either half-space or the exterior of a sphere (but the half-space setting is used only occasionally for localized regional distributions, so it is not stressed).
\item 
 The study of kernels is broken down into the standard empirically modeled statistical covariance kernels and analytical collocation (where closed-form kernels are studied)---analytical collocation is generally only used when the functional form of the kernel is understood to be a workable representation of the statistical covariance function.
\item 
  Field measurement errors are allowed.
\end{enumerate}
  Practitioners of GC have cataloged most, if not all, useful kernel possibilities allowed by these five differences.  The main interpretational basis of GC is the GC or minimum norm property previously mentioned in connection with the GLLSQC condition in Section~\ref{S:GLLSQ}:  Under the assumption of errorless measurements, from all possible candidate functions that reproduce the given point measurements, GC selects the one that is smoothest---that is the one with the smallest norm, where the norm is determined by the underlying covariance function \cite[pp. 207--220]{Moritz}.  It this minimum norm property, in concert with the emphasis on statistical covariance functions indicated by the first point above, that makes GC approaches less sensitive to kernel point  placement issues; however, as indicated above this also implies a corresponding loss of fitting responsiveness and thus, for example, it would be hard to argue that GC is capable of the overall level of economy and efficency indicated above for the NLLSQ DIDACKS low degree and order tesseral fits.  Further observe that just as in RKHS theory, GC also satisfies a least squares norm property \cite{Moritz}, but in practice this property is usually ignored by physical geodesists.

  For LLSQ gravity reconstruction problems GC has historically been the first procedure of choice and for these problems one could argue that when good covariance data is available, these techniques are safe and easy to use; however, as classically practiced GC techniques do have notable limitations, which are not shared by the DIDACKS approach:
\renewcommand{\theenumi}{\Alph{enumi}}  
\renewcommand{\labelenumi}{(\theenumi)} 
\begin{enumerate}
\item 
 NLLSQ applications cannot be treated.
\item 
 GC techniques are customarily applicable only to the Earth's gravitational field environment since they require covariance information that is often only available in this context.
\item 
 The application of GC requires a certain level of familiarity and thus it is almost never adapted to problems outside the geophysical realm, even when appropriate covariance data can be gathered.  (There are, however, various other areas that use somewhat related kernel techniques \cite{SandW}.)
\item 
 Inverse source estimation problems cannot be entertained.
\end{enumerate}

    In summary, for gravity reconstruction problems with accurate selected data sets where either approach can be used, results using the DIDACKS approach can be both much better or much worse than one might normally expect with GC approaches since the ingenuity, implementation skill and patience of the practitioner have a much more significant bearing on the outcome for DIDACKS approaches.  GC is and probably will always remain the primary mathematical tool employed for raw gravity data processing and related uses where efficiency is not the primary concern, since its behavior in these arenas is well understood and it can handle measurement errors naturally.

\section{Mathematical Connections to Geophysical Collocation}\label{S:GCconnections}

  As indicated in Section~\ref{S:intro}, (\ref{E:mkeqn2}a) can be related to a line of preexisting research performed by Krarup in conjunction with his studies of GC \cite{Krarup}.  Also an independent parallel line of point source research exists that maintains direct connections to GC itself and can be considered a direct outgrowth of Krarup's original work.  This alternative collocation based point source scheme is briefly considered after the connections of DIDACKS theory to Krarup's work are addressed.  Since this alternative scheme and Krarup's work are primarily based on GC for spherical exteriors, only this geometry will be considered in the sequel.

  Krarup first introduced the weighted integral occurring on the RHS of (\ref{E:inorm2d}) in conjuction with his study of GC \cite[pp. 62--65]{Krarup}.  The SRK corresponding to this norm (the ``Krarup kernel,'' $K_{\smallindex{K}}$) has the form
\begin{equation}\label{E:KrarupKernel}
K_{\smallindex{K}}(\vec{P},\,\vec{Q}) = \frac{R_{\sps}}{\sqrt{R_{\sps}^2 - \, 2\,\vec{P}\!\cdot\vec{Q}\, +\, \left({P\,Q}/{R_{\sps}}\right)^2}}\ ,
\end{equation}
which Krarup and his followers extensively studied and applied.  In (\ref{E:KrarupKernel}) $Q = |\vec{Q}|$.  To understand how point mass fitting enters, observe that this SRK can be recast in the form $R_{\sps}^2/(P|\vec{Q} - \vec{X}'|)$, or $|\vec{X}'|/|\vec{Q} - \vec{X}'|$ since $\vec{X}' = \vec{P}R_{\sps}^2/P^2$, which is proportional to a point mass potential at a fixed location.  Making the ansatz $\vec{X}' \longrightarrow \vec{X}'_k$ and $\vec{Q} \longrightarrow {\vec{X}}$ allows one to transform a collocation fit into a point mass fit if the mass locations are restricted to be at a fixed depth $r'_k = |\vec{X}'_k| =$ constant, where $r'_k$ is simply treated as an overall constant of proportionality.  The relevant history of this and related ideas is briefly touched on below.  Here one possible next step, which Krarup and his followers did not apparently take since it entails abandoning symmetric kernel forms altogether, is to generalize this procedure to independent variable depths by absorbing the factors ${P}_k \eq R_{\sps}^2/r'_k$ and $R_{\sps}$ into each of the collocation fitting parameters separately and then reinterpreting the resulting collocation fit as a point mass fit, which yields a fit based on (\ref{E:mkeqn2}a) as the end product.

   Conversely, DIDACKS point mass fits based on the integral norm can be reinterpreted as collocation fits \cite{ruf}.  The resulting fits are based on what is called the reciprocal distance covariance function \cite[p. 182]{Moritz}, which corresponds to $R^2_{\sps}/PQ$ times the Krarup kernel specified by (\ref{E:KrarupKernel}).  Notice that while the integral norm can be reexpressed in terms of the ``Krarup norm'' by (\ref{E:inorm2d}), it is useful to retain the integral norm as a distinct entity specified by (\ref{E:intnorm}) since (a) it is a surface integral rather than a volume integral, (b) there are some consequences of this form, such as (\ref{E:rel2}), that are not apparent from the weighted Dirichlet integral form itself and (c) the Krarup norm is associated exclusively with the Krarup kernel in SRK form and it is primarily linked to GC theory where, as previously noted, the goals and practices are quite different.
 
   As summarized in \cite{MarchenkoLelgemann} and as just indicated, starting in the early 1980's a distinct line of point source research based on Krarup's original work above was developed by A. N. Marchenko and others \cite{HauckLelgemann}.  This research is based on maintaining connections to symmetric global GC covariance kernel forms.  For the spherical case, assuming the usual GC covariance properties, the general form of allowed global covariance kernel, $C(\vec{P},\,\vec{Q})$, can be written as:
\begin{equation}\label{E:GCkernel}
C(\vec{P},\,\vec{Q})\, =\, \sum\limits_{n=0}^{\infty}\, k_n \left(\frac{\ R_{\sps}^2}{PQ}\right)^{n+1}{\text{P}}_n(\cos \psi)\,, \ \ \text{where}\ \ \cos \psi\, = \,(\vec{P}\!\cdot\!\vec{Q}/P\,Q)\,,
\end{equation}
where the ${\text{P}}_n$ are standard (un-normalized) Legendre polynomials and the $k_n$ are constants \cite[p. 181]{Moritz}. Here, as before, $R_{\sps}$ is the radius of the spherical region.  Mathematically, a kernel specified by (\ref{E:GCkernel}) can, in general, be simply considered a SRK with an added layer of statistical interpretation.  Solving (\ref{E:mkeqn2}b) for ${\vec{X}}' = {\vec{X}}'(\vec{P})$ and substituting the result into (\ref{E:GCkernel}) yields a kernel that is a function of one interior point and one exterior point.  For certain applications, when (\ref{E:GCkernel}) is reexpressed in terms of a smaller radius $R_B < R_{\sps}$ the resulting sphere is known as a Bjerhammar sphere.  For particular choices of $k_n$ if (\ref{E:GCkernel}) can be rewritten as a closed-form expression and if this expression has the correct form, such as (\ref{E:KrarupKernel}), then the result can be reexpressed as a point mass or other linear combinations of point or line source potentials.  As just noted in connection with (\ref{E:KrarupKernel}), additional parameters will occur in the resulting expressions that do not appear in the source potentials themselves, but if the locations of the sources are restricted these parameters can be explained away as common constant factors.  A Bjerhammar sphere is generally used since it allows for an independent adjustment of the overall depth of the interior source points.  By maintaining connections to the symmetric kernel forms that are allowed by (\ref{E:GCkernel}), a statistical covariance interpretation is possible. 

  As one might expect there are notable differences between this approach and DIDACKS since this alternative approach adopts the basic philosophy of maintaining connections to GC and DIDACKS does not.  In particular for a specified domain, from RKHS theory one can deduce that the choice of inner product (or norm) and reproducing kernel must be in one-to-one correspondence \cite{Moritz}; hence, the choice of multipole form to be fit and norm are directly linked.  A list of known covariance kernel and multipole point source correspondences for this approach can be found in \cite{MarchenkoLelgemann}.  For example, in \cite{HauckLelgemann} some of the consequences of the Krarup norm choice, along with its half-space approximation, are examined for point mass fits.  By employing (\ref{E:GCkernel}) this approach allows for an underpinning statistical covariance interpretation, which is important for many geophysical applications; nevertheless, implicitly retaining symmetric kernel forms also adds a layer of additional complexity, which has the effect of greatly complicating the theory and of limiting the possible choices of dipole and higher multipole forms.  For uses outside geophysics there are also clearly interpretational difficulties that limit its use.  

 In DIDACKS theory all attempts at retaining connections to SRKs are dropped and the primary emphasis is placed on the (weighted) energy or integral norm and the associated kernel form $\ell^{-1}$.  As noted earlier, this makes consideration of dipoles and higher multipoles trivial since the required inner products are readily obtained. Moreover, since the DIDACKS approach maintains the same norm choice for all source types there are few interpretational issues---especially with regards to multipole fits of all orders.  The types of geophysical areas where one of these two approaches or GC should be preferred over the others clearly warrants further study and consideration since, to date, there have been no researchers proficient in applying all three algorithms.  (While GC has had many practitioners and this alternative collocation based approach has had a few practitioners, so far DIDACKS development and use has been limited to the author's involvement.)




\end{document}